\documentclass[prb,twocolumn, showpacs, groupedaddress]{revtex4}

\usepackage{graphicx}
\usepackage{dcolumn}
\usepackage{bm}

\begin{document}

\preprint{LMH-1}

\title{Spin gaps and magnetic structure of Na$_{x}$CoO$_2$}

\author{L. M. Helme}
\email[Electronic address: ]{l.helme1@physics.ox.ac.uk}
\author{A. T. Boothroyd}
\author{R. Coldea}
\author{D. Prabhakaran}
\affiliation{Department of Physics, University of Oxford, Oxford,
OX1 3PU, United Kingdom }

\author{A. Stunault}
\author{G. J. McIntyre}
\author{N. Kernavanois}
\affiliation{ Institut Laue-Langevin, BP 156, 38042 Grenoble Cedex
9, France }

\date{\today}

\begin{abstract}

We present two experiments that provide information on spin
anisotropy and the magnetic structure of Na$_{x}$CoO$_2$. First,
we report low-energy neutron inelastic scattering measurements of
the zone-center magnetic excitations in the magnetically ordered
phase of Na$_{0.75}$CoO$_2$. The energy spectra suggest the
existence of two gaps, and are very well fitted by a spin-wave
model with both in-plane and out-of-plane anisotropy terms. The
gap energies decrease with increasing temperature and both gaps
are found to have closed when the temperature exceeds the magnetic
ordering temperature $T_m \approx 22$~K. Secondly, we present
neutron diffraction studies of Na$_{0.85}$CoO$_2$ with a magnetic
field applied approximately parallel to the $c$ axis. For fields
in excess of $\sim 8$~T a magnetic Bragg peak was observed at the
$(0,0,3)$ position in reciprocal space. We interpret this as a
spin-flop transition of the A-type antiferromagnetic structure,
and we show that the spin-flop field is consistent with the size
of the anisotropy gap.

\end{abstract}

\maketitle

\section{Introduction} 

Since the discovery of superconductivity in water-intercalated
Na$_x$CoO$_2$ \cite{takada-2003} much of the discussion has
focussed on the mechanism of superconductivity. There is strong
support from experiment \cite{experimental-evidence} and theory
\cite{mazin-2005, theoretical-evidence} for an unconventional
pairing state, the origin of which derives from the triangular
lattice of the Co ions and the existence of strong spin and charge
fluctuations. However, details of the pairing state are far from
resolved. \cite{mazin-2005} The phase diagram of Na$_x$CoO$_2$
shows two magnetically ordered phases, one at $x \approx 0.5$ and
the other in the range $x\approx 0.7 - 0.95$. The presence of
these suggests that magnetic correlations may play an important
role in the formation of superconductivity in hydrated
Na$_x$CoO$_2$, as is believed to be the case for the
superconducting cuprates. This possibility provides a strong
incentive for characterizing the magnetic order and excitations of
Na$_x$CoO$_2$.

\begin{figure}[!ht]
\begin{center}
\includegraphics{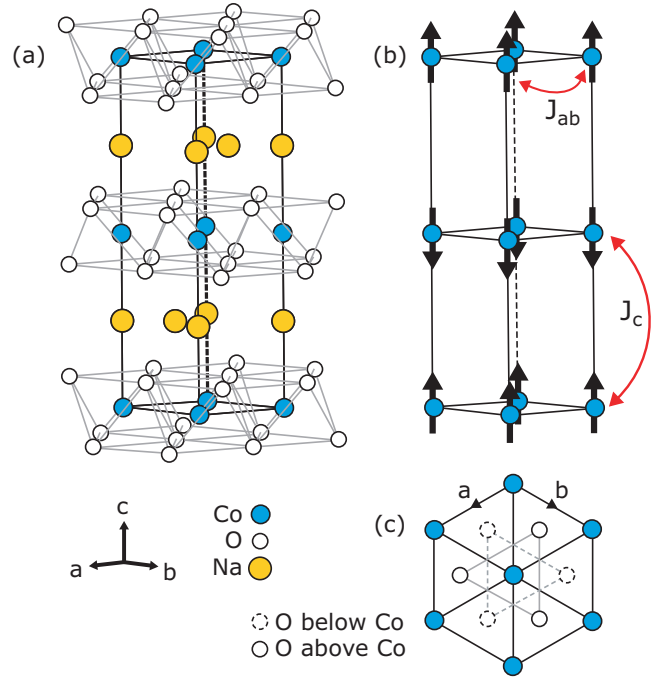}
\caption[]{(color online) (a) Crystal structure of Na$_x$CoO$_2$.
(b) The A-type antiferromagnetic structure on which the spin-wave
model is based, showing the two exchange constants $J_{ab}$ and
$J_c$. (c) The $a$--$b$ planes, showing the orientation of the
oxygen tetrahedron around a central Co ion.} \label{fig:structure}
\end{center}
\end{figure}

This paper is concerned with the weakly magnetic phase found for
$x\approx 0.7 - 0.95$ at temperatures below $T_m \approx
22$~K.\cite{Motohashi-2003,Sugiyama-2004} Recent neutron
scattering studies of Na$_{0.75}$CoO$_2$
\cite{boothroyd-may04,helme-2005} and Na$_{0.82}$CoO$_2$
\cite{bayrakci-2005} have established that the magnetic order and
dynamics are consistent with an A-type antiferromagnetic
structure, shown in Fig.~\ref{fig:structure}, and that the
magnetic interactions are three-dimensional despite the
two-dimensional character of the crystal lattice and electronic
structure. Reasons for the relatively strong $c$-axis magnetic
coupling have been discussed recently by Johannes {\it et
al.}\cite{Johannes-2005} Surprisingly, although the magnetic
correlations are rather strong the ordered moment is only
0.1--0.2~$\mu_{\rm B}$.\cite{Sugiyama-2003,bayrakci-2005}

The first aim of this work was to extend the measurements of the
spin-wave dispersion in Na$_{0.75}$CoO$_2$ down to lower energies
where we might gain important information on the magnetic ground
state, such as whether itinerant effects are important. Our
previous measurements,\cite{helme-2005} as well as those by
Bayrakci {\it et al.} on Na$_{0.82}$CoO$_2$,\cite{bayrakci-2005}
found evidence of a small excitation gap at the antiferromagnetic
zone center. Here we present a detailed study of the spectral
lineshape in this low-energy region which not only confirms that
the magnetic excitation spectrum is gapped, but also provides
evidence for the existence of two gaps. Both these gaps are found
to close at the bulk magnetic transition temperature $T_m \approx
$22~K.

The second part of the paper presents a novel way to investigate
the magnetic structure of Na$_x$CoO$_2$ ($x \approx 0.7-0.95$) and
also to gain information about spin anisotropy. In the accepted
A-type antiferromagnetic structure (Fig.~\ref{fig:structure}b) the
moments are ferromagnetically aligned within the layers and
stacked antiferromagnetically along the $c$
axis.\cite{helme-2005,bayrakci-2005} The ordering wavevector for
this structure is $(0,0,1)$.\footnote{We label wavevectors, ${\bf
Q}$, in terms of the reciprocal lattice units $(h, k, l)$ of the
hexagonal unit cell (Fig.\ref{fig:structure}).} Strong spin-wave
scattering is observed emerging from $(0,0,l)$ positions with odd
$l$, which are zone centers for the A-type antiferromagnetic
order, but no magnetic Bragg peaks have been observed at these
positions using neutrons. \footnote{There is strong spin-wave
scattering around ${\bf Q}=(0,0,l)$ with $l=$~odd because the spin
fluctuations are perpendicular to ${\bf Q}$.} On this basis it was
deduced that the ordered moments point along the $c$ direction
since neutrons scatter from the component of the moments
perpendicular to the scattering vector. This moment direction is
consistent with what has been inferred from the uniform
susceptibility and from $\mu$SR data.\cite{moment-reference}
Moreover, Bayrakci {\it et al.} did succeed recently in observing
magnetic Bragg reflections at a few $(h,k,l)$ positions with $h$,
$k\neq 0$ and odd $l$ using polarized
neutrons.\cite{bayrakci-2005} Again, these are consistent with the
accepted magnetic structure. Polarized neutrons were required
because the ordered moment is small and strong non-magnetic
scattering is observed at all positions where magnetic Bragg peaks
are expected. Up to now no magnetic Bragg peaks have been observed
with unpolarized neutrons.

The experiment we describe here was originally designed to confirm
the proposed magnetic structure by a method that employs
unpolarized neutrons but avoids the problem of having to separate
the weak magnetic scattering from the strong non-magnetic
background signal. Our approach was motivated by measurements of
the magnetization of Na$_{0.85}$CoO$_2$ in applied fields up to
14~T by Luo {\it et al.}. \cite{luo-2004} The magnetization data
with $H \parallel c$ show a clear anomaly at 8~T at low
temperatures, with no such transition seen for $H \perp c$. The
authors interpreted this as a spin-flop transition in which the
ordered moments rotate by $\sim$90 degrees while preserving the
A-type antiferromagnetic arrangement. After the transition the
spins lie approximately in the hexagonal plane, but the magnetic
structure has a small ferromagnetic component along the $c$ axis.
Assuming this explanation to be correct we induced the spin-flop
transition in a neutron diffraction experiment and searched for
magnetic Bragg peaks along ${\bf Q}=(0,0,l)$, since now the
ordered moment should be perpendicular to the scattering
wavevector and should scatter neutrons. Our experimental results
are in excellent accord with the predicted behavior. Encouraged by
this we go on to show that the size of the spin gap in the
magnetic excitation spectrum is in agreement with the observed
spin-flop field. This quantitative analysis provides the link
between the static and dynamic magnetic properties explored in
this paper.

The remainder of the paper is organized as follows. Inelastic
neutron measurements probing the low-energy region of the
spin-wave dispersion are presented in the next section, along with
the analysis of these results. Diffraction measurements made with
an external magnetic field applied parallel to the $c$ axis are
presented and analyzed in Sec. III. Section IV contains a
discussion of the results, and the conclusions are presented in
Sec. V.

\section{Inelastic Measurements}

\subsection{Experimental Details}

Inelastic neutron measurements were performed on the cold-neutron
triple-axis spectrometer IN14 at the Institut Laue-Langevin. This
instrument was chosen to allow investigation of the excitations in
Na$_{0.75}$CoO$_2$ at lower energies than previously
studied.\cite{helme-2005} We employed a pyrolytic graphite (PG)
$(002)$ monochromator and a PG $(002)$ analyzer, which were curved
vertically and horizontally respectively to maximize the count
rate. The majority of measurements were made with a fixed final
energy of $E_{\rm f}$ = 4~meV. A Beryllium filter was placed in
the scattered beam to suppress higher-order harmonics.

The inelastic neutron measurements were performed on the same
crystal of Na$_{0.75}$CoO$_2$ as used for our previous experiments
which investigated the spin fluctuations in this compound at
higher energies.\cite{helme-2005} The single crystal of mass $\sim
1.5$~g was mounted on a copper mount and aligned to allow
measurements to be made within the $(100)$--$(001)$ scattering
plane. Details of the growth and mounting are given elsewhere.
\cite{helme-2005,prabhaks-paper}

Previous measurements revealed strong spin-wave scattering around
(0,0,1) and (0,0,3), consistent with the proposed A-type
antiferromagnetic structure, and the inelastic data presented here
concentrates on spin waves dispersing from the magnetic zone
center at (0,0,1), where the inelastic scattering is most intense.

\subsection{Results}

\begin{figure}[!t]
\begin{center}
\includegraphics{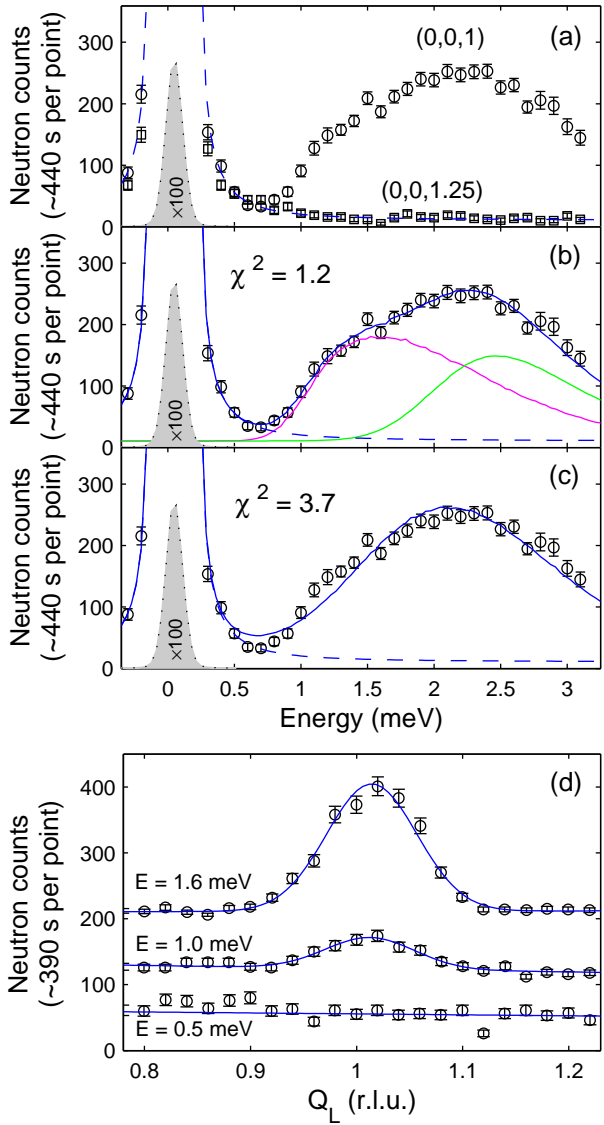}
\caption[Two modes]{(color online) Neutron inelastic scattering
from Na$_{0.75}$CoO$_2$ measured at $T=1.5$~K. (a) Energy scan at
constant ${\bf Q}=(0, 0, 1)$, compared with the same scan at a
background position ${\bf Q}=(0, 0, 1.25)$. (b, c): The data at
${\bf Q}=(0, 0, 1)$, fitted (solid line) with (b) a two-mode
dispersion (Eqn.\ref{dispersion-relation}), and (c) one mode only
(Eqn.\ref{dispersion-relation} but with $E\equiv0$). The two peaks
under the data in (b) show the contribution of each mode to the
total intensity, while in (a-c) the dashed curve represents the
contribution of the incoherent peak and background. The shaded
peaks show the incoherent peak contribution scaled down by a
factor of 100, as an indication of the instrumental resolution.
(d) ${\bf Q}_l$ scans with constant energy transfers of 0.5~meV,
1.0~meV and 1.6~meV. Data at 1.0~meV and 1.6~meV have been shifted
up by 100 and 200 counts respectively for clarity.}
\label{fig:two_modes}
\end{center}
\end{figure}

\begin{figure}[!ht]
\begin{center}
\includegraphics{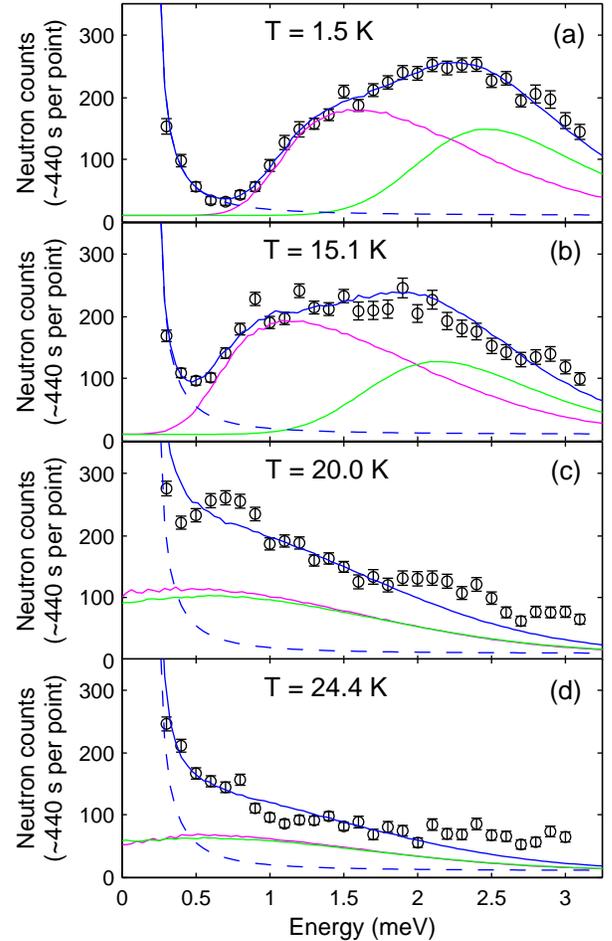}
\caption[]{(color online) (a)-(d) Energy scans at ${\bf Q}=(0, 0,
1)$ at temperatures between $T=1.5$~K and 24.4~K. Solid curves
represent the best fit of Eqn.\ref{dispersion-relation} plus an
incoherent peak to the data. The dashed line shows the
contribution of the incoherent peak and background, while the pink
and green curves represent the intensities of each mode.}
\label{fig:temp_dep}
\end{center}
\end{figure}

Figures \ref{fig:two_modes} and \ref{fig:temp_dep} present
examples of inelastic neutron scattering data collected on IN14.
Each scan was performed by measuring the intensity of scattered
neutrons as a function of energy transfer up to $\sim 3$~meV at
the wavevector ${\bf Q}=(0, 0, 1)$. Scans were made at ten
temperatures between 1.5~K and 24.4~K.

Figures \ref{fig:two_modes}a--c show the measurements made at low
temperature ($T=1.5$~K). The spectrum consists of an intense peak
due to incoherent nuclear elastic scattering centered on
$E=0$~meV, and a broad signal centered around 2~meV which is
attributed to magnetic scattering as the scan cuts through the
spin-wave dispersion. There is clearly a gap where the intensity
falls to background below $\sim 1$~meV, revealing that the
magnetic excitations in Na$_{0.75}$CoO$_2$ are separated from the
ordered ground state by a clean gap.
%
To determine the non-magnetic scattering an energy scan was also
made at ${\bf Q}=(0, 0, 1.25)$. The scan, which is plotted in
Fig.~\ref{fig:two_modes}a, contains the nuclear incoherent peak
together with a small constant background signal.
Figure~\ref{fig:two_modes}d displays ${\bf Q}_l$ scans performed
at three constant energy transfers of $E_T = 0.5$, 1.0 and
1.6~meV. The peak present at higher energies has clearly
disappeared at 0.5~meV, confirming that the intensity of the
spin-wave dispersion really does fall to background in the `gap',
and the remaining intensity at this point seen in
Fig.s~\ref{fig:two_modes}a--c is simply due to the tail of the
incoherent peak.

Figure \ref{fig:temp_dep} shows the same scan made at three more
of the temperatures we measured, below and above the magnetic
transition temperature $T_m \approx 22$~K. It appears that the
magnetic scattering intensity moves lower in energy as the
temperature increases. Somewhere around 20~K the gap seems to
disappear, moving into the incoherent peak.

\subsection{Analysis}

To determine whether the gap is in fact decreasing with
temperature, or whether what we see is simply the mode broadening
and decreasing in intensity, it is necessary to compare the
experimental results with a model of the excitations. The scan
observed at 1.5~K (Fig.\ref{fig:two_modes}a) is suggestive of a
two-peak lineshape. In order to extend the spin-wave model
introduced in our previous work \cite{helme-2005} to allow two
non-degenerate gapped modes at ${\bf Q}=(0, 0, 1)$ we include two
anisotropy terms. The Hamiltonian of this refined model is then:
\begin{eqnarray}
    \mathcal{H}&=&J_{ab}\sum_{\langle i, i^\prime \rangle} {\bf S}_i
    \cdot {\bf S}_{i^\prime} + J_c\sum_{\langle i, j \rangle}  {\bf S}_i
    \cdot {\bf S}_j \nonumber\\
                &   & - D\sum_{i} (S_i^z)^2 - E\sum_{i} [(S_i^x)^2 - (S_i^y)^2]\;\; , \label{hamiltonian}
\end{eqnarray}
where $J_{ab}$ and $J_c$ are intra- and inter-layer exchange
parameters, respectively, as in our previous work.
\cite{helme-2005} The anisotropy constant $D$ quantifies the
tendency of the spins to lie along the $c$ axis, \footnote{We
should note that Bayrakci {\it et al.} introduced a similar term
$- D\sum_{i} (S_i^z)$ (with the sign of $D$ alternating from layer
to layer) to describe a single anisotropy gap, in their paper on
Na$_{0.82}$CoO$_2$. However they were unable to determine
definitively the existence of the gap, fitting a value for $|D|$
of $0.05\pm 0.05$. \cite{bayrakci-2005}} while the term $E\sum_{i}
[(S_i^x)^2 - (S_i^y)^2]$, which has two-fold symmetry in the
plane, is the simplest way to introduce in-plane anisotropy. We
define $x$ parallel to $a$, $z$ parallel to $c$ and $y$
perpendicular to $x$ and $z$ so as to make a right-handed set.
From the hexagonal arrangement of the Co ions within the $a$--$b$
layers of this compound one might propose a term with hexagonal
symmetry for the in-plane anisotropy. However, with the
quantization direction parallel to the $c$ axis a term with
hexagonal symmetry does not lift the two-fold degeneracy of the
spin-wave dispersion, \footnote{In fact any term containing only
products of $S^x$ and $S^y$ higher than order two will not
generate a gap when the spins lie along the $z$-direction.} so for
the purposes of this analysis we use the Hamiltonian above (Eqn.
\ref{hamiltonian}).

The spin-wave dispersion resulting from this Hamiltonian can be
calculated using standard methods, giving two modes:
\begin{equation}\label{dispersion-relation}
\hbar\omega^\pm_{\mathbf Q}=2S\sqrt{(A_\mathbf{Q}+D)^2 -
(C_\mathbf{Q} \pm E)^2} \;\; ,
\end{equation}
where $\hbar\omega$ is the energy transfer, $S$ is the spin (here
assumed to be $S$ = 1/2), ${\mathbf Q}=(h,k,l)$ is the wavevector,
and $A_{\mathbf Q}$ and $C_{\mathbf Q}$ are defined in terms of
the exchange couplings as
\begin{eqnarray}\label{A_Q_C_Q}
    A_{\mathbf Q}&=&J_{ab}\{\cos(2\pi h) + \cos(2\pi k)+ \cos[2\pi (h + k)]-3\}  \nonumber \\
                 & &+J_c \nonumber \\
    C_{\mathbf Q}&=&J_c \cos(\pi l)
    .\;\;
\end{eqnarray}
The magnitude of the gaps at the magnetic zone center are then
related to the exchange and anisotropy parameters as follows:
\begin{equation}\label{gap_energies}
    \hbar\omega_{gap}^\pm =  2S\sqrt{(J_c + D)^2 - (J_c \pm E)^2} \;.
\end{equation}
Note that if $E=0$ only one gap results.


For the case when ${\mathbf Q}$ lies parallel to the ordered
moment direction, such as at ${\mathbf Q}=(0,0,1)$ here, the
inelastic neutron intensity is proportional to $S^{xx}(\mathbf{Q},
\omega)+S^{yy}(\mathbf{Q}, \omega)$, \cite{squires} where
\begin{eqnarray}
S^{xx}(\mathbf{Q},\omega) & = & 2
S^2\frac{\{(A_{\mathbf{Q}}+D)-(C_{\mathbf{Q}}-E)\}}{\hbar\omega^-_{\mathbf{Q}}}
\nonumber \\
&  & \; \times G^-(\omega-\omega^-_{\mathbf{Q}}) f^2(\mathbf{Q})[n(\omega)+1]       \label{Sxx} \\
 S^{yy}(\mathbf{Q},\omega) &= & 2 S^2
\frac{\{(A_{\mathbf{Q}}+D)-(C_{\mathbf{Q}}+E)\}}{\hbar\omega^+_{\mathbf{Q}}}\nonumber \\
&  & \; \times G^+(\omega-\omega^+_{\mathbf{Q}})
f^2(\mathbf{Q})[n(\omega)+1] \;\; ,\label{Syy}
\end{eqnarray}
where $[n(\omega)+1]$ is the Bose factor, and $f(\mathbf{Q})$ is
the form factor (which is a constant here since all our energy
scans were made at one fixed value of $\mathbf{Q}$).
$G(\omega-\omega^{\pm}_{\mathbf{Q}})$ are normalized Gaussian
functions which replace the usual Delta functions to allow
inclusion of intrinsic broadening of the two modes.

The triple-axis spectrometer has a three-dimensional
ellipsoid-shaped resolution, and therefore does not probe the
dispersion relation at an infinitely sharp point in reciprocal
space. In order to fit the spin-wave model to the experimental
data it was therefore necessary to convolute the calculated
spectrum (Eqns.~\ref{Sxx} and \ref{Syy}) with the IN14
spectrometer resolution.
 This was achieved using RESCAL, a set of
programs integrated into Matlab which calculates the resolution
function of the neutron triple-axis spectrometer.\cite{rescal} It
allows simulation of scans using a 4D Monte-Carlo convolution of
the resolution function with the specified spectrum, and the
simulation can then be fitted to the data in order to extract the
parameters.

In this way the anisotropy parameters were extracted, while fixing
the exchange parameters $J_{ab}$ and $J_c$ to their previous
values of $-6$~meV and 12.2~meV respectively. \cite{helme-2005}
The relative amplitudes of the two modes were fixed by the
spin-wave model, with an overall amplitude fitted, the values for
the intrinsic widths of the dispersion modes were fitted
independently. The incoherent peak and background were included as
a fixed Voigtian peak plus a constant.

At $T=1.5$~K the values for $D$ and $E$ were found to be $0.096
\pm 0.005$ and $0.059 \pm 0.005$~meV, corresponding to two modes
with gaps of $0.95 \pm 0.13$ and $1.95 \pm 0.06$~meV. To achieve a
good fit the intrinsic widths of the two modes were found to be
different, 0.37 and 0.74~meV for the lower and higher modes. The
fitted curve is displayed on Fig.\ref{fig:two_modes}a, with the
two lower curves representing the contribution of each of the
gapped modes. The dashed line shows the contribution of the
incoherent peak, which is also plotted scaled down by a factor of
100 (shaded peak) as an indication of the instrumental resolution.

\begin{figure}[!ht]
\begin{center}
\includegraphics{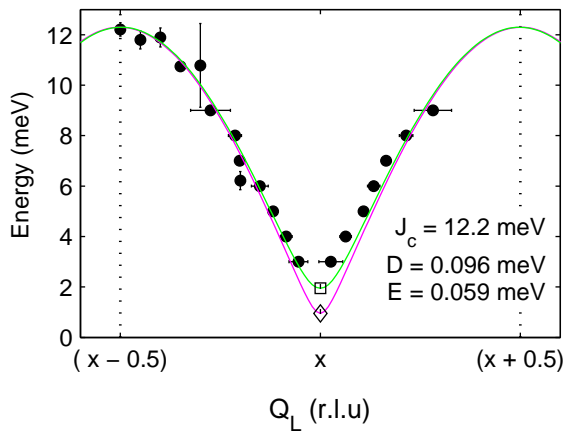}
\caption[]{(color online) Filled circles: Magnon dispersion
parallel to $(0,0,l)$ centered on $(0,0,x)$=$(0,0,3)$, measured
previously.\cite{helme-2005} Open square/diamond: Fits to 1.5~K
energy scan at $(0,0,x)$=$(0,0,1)$ shown in
Fig.\ref{fig:two_modes}. Solid curves are modes calculated from
the spin-wave dispersion Eqn.\ref{dispersion-relation} with the
constants given, while dotted lines represent the zone
boundaries.} \label{fig:overplotted}
\end{center}
\end{figure}

The model appears to fit the data well. For comparison
Fig.~\ref{fig:two_modes}b shows the same data fitted with a
`one-mode' dispersion, by fixing the value of $E$ to zero. It is
clear that the model with two modes fits the data better than that
with one, yielding a value of $\chi^2$=1.2 compared to
$\chi^2$=3.7 with $E=0$. In Fig.~\ref{fig:overplotted} we plot the
two dispersion modes parallel to ${\bf Q}_l$ (calculated from
Eqn.~\ref{dispersion-relation} using the fitted parameters for $D$
and $E$), together with the data previously measured around ${\bf
Q}=(0,0,3)$. The fitted modes are also in good agreement with the
experimental data in the ${\bf Q}_l$ direction.

\begin{figure}[]
\begin{center}
\includegraphics{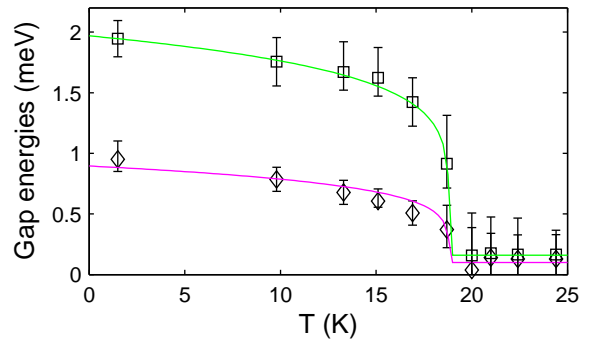}
\caption[]{(color online) The magnitudes of the two gaps as a
function of temperature. Data points calculated from
Eqn.\ref{gap_energies} using the values of $J_c$, $D$ and $E$
derived from fits such as those in Fig.\ref{fig:temp_dep}. Solid
curves are guides for the eye. Error bars were estimated by
varying the two gap energies separately until the fit was no
longer acceptable.} \label{fig:gap_params}
\end{center}
\end{figure}

The fitting procedure was repeated for data at all temperatures,
restricted only by fixing the intrinsic widths of the two modes to
the values at $1.5$~K. Figures~\ref{fig:temp_dep}a--d are
overplotted with fits to each data set, with the contributions
from each mode as solid lines underneath, and the contribution
from the incoherent peak denoted by the dashed line. The fitted
lines provide a reasonable description of the data, reproducing
the shift of the intensity towards zero energy with increasing
temperature, although clearly the lineshapes fit less well as the
temperature increases. The gap energies extracted from these fits
are plotted as a function of temperature in Fig.\
\ref{fig:gap_params}.

Both gaps are seen to decrease with temperature, falling to near
zero at $\approx 20$~K. Above $20$~K the fitted gaps are relatively
constant and close to zero.

\section{Diffraction Measurements}

\subsection{Experimental Details}

Neutron diffraction measurements were performed on the hot-neutron
diffractometer D3 at the Institut Laue-Langevin. The instrument
was used in unpolarized-neutron mode with a neutron wavelength of
0.84~{\AA}. The single crystal of Na$_{0.85}$CoO$_2$ used for
these measurements was cleaved from a rod grown in Oxford by the
floating-zone method.\cite{prabhaks-paper} The crystal had a mass
of 0.3~g and a mosaic spread of $\sim 2$~degrees.

Magnetization measurements made using a SQUID magnetometer in
Oxford confirmed the existence of the metamagnetic transition at
$\sim 9$~T reported by Luo {\it et al.} at this composition.
\cite{luo-2004}

The crystal was pre-aligned on the neutron Laue diffractometer
Orient Express at the ILL and mounted on an aluminium pin using
ceramic glue. The ideal setup for this experiment would be to
align the $c$ axis vertically, applying a vertical field, with the
incident and scattered beams inclined at the Bragg angle to the
horizontal in order to access the $(0,0,3)$ reflection. An
alternative setup, with a fixed horizontal incident beam, would
require the cryomagnet holding the sample to be tilted by the
Bragg angle ($\theta_{B}$), with the detector lifted out of the
plane (by $2\theta_{B}$), allowing access to the $(0,0,3)$
reflection while still applying the field directly along the $c$
axis.

However, on D3 we were restricted to using a horizontal incident
beam (ruling out the first setup), and also unable to tilt the
cryomagnet (ruling out the second). In order to access the
$(0,0,3)$ reflection we tilted the crystal $c$ axis $7$ degrees
away from vertical, corresponding to the $(0,0,3)$ Bragg angle for
0.84~{\AA} neutrons, and lifted the detector out of the horizontal
plane by 14~deg. The 10~Tesla vertical-field cryomagnet in which
the crystal was mounted then allowed application of a field almost
parallel to the $c$ axis (though actually $7$ degrees away from
it). The field-induced (0,0,1) magnetic Bragg peak is expected to
be larger than that at (0,0,3) due to the magnetic form factor,
but with 0.84~{\AA} neutrons the scattering angle for (0,0,1) is
too small to access with our setup.

\subsection{Results}

\begin{figure}[!t]
\begin{center}
\includegraphics{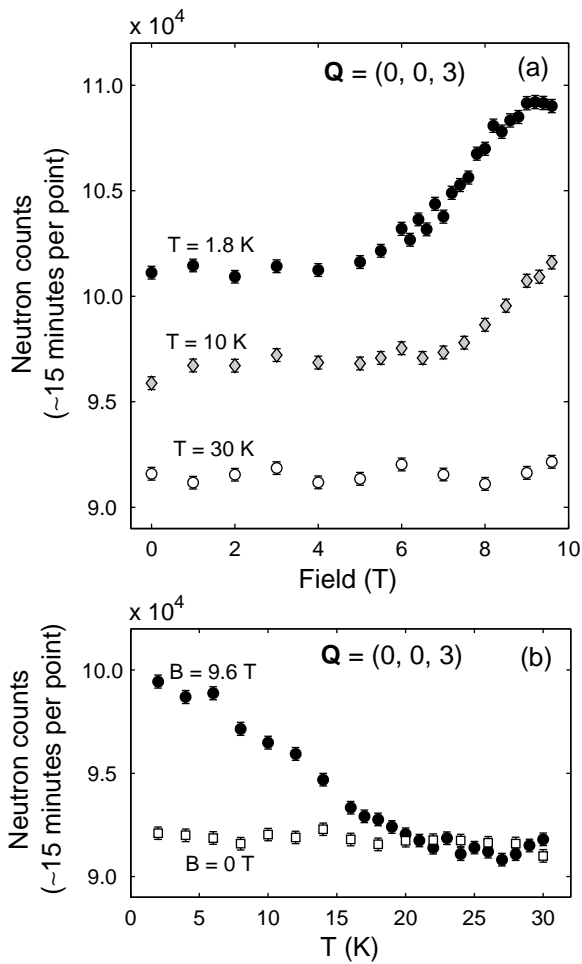}
\caption[Caption Caption]{Diffraction studies of
Na$_{0.85}$CoO$_2$. (a) Field scans at ${\bf Q}=(0, 0, 3)$ at
constant temperatures of $T=1.8$, 10 and 30~K. Data at $T=10$~K
and 30~K are shifted up by 500 and 1000 counts respectively. (b)
Temperature scans at ${\bf Q}=(0, 0, 3)$, with zero applied field
(open circles), and with $H=$9.6~T applied at 7 degrees to the
$c$-axis (filled circles). } \label{fig:D3}
\end{center}
\end{figure}

Figure \ref{fig:D3} shows the main results of the diffraction
studies of Na$_{0.85}$CoO$_2$. All the measurements were made at
${\bf Q}=(0, 0, 3)$, scanning either field or temperature with the
other external variable fixed. For ideal Na$_x$CoO$_2$ no
structural Bragg peak is allowed at this position and no magnetic
Bragg peak is allowed if the moments point along the $c$ axis.

Figure \ref{fig:D3}a shows three field scans at ${\bf Q}=(0, 0, 3)$,
performed at constant temperatures of $T=1.8$, 10 and 30~K. The data
at 10~K and 1.8~K have been shifted up by 500 and 1000 counts
respectively for clarity.

At $T=1.8$~K there is a large increase in the intensity of
scattering at $(0,0,3)$ between $\sim$6~T and 9~T as the field
increases, with the intensity appearing to flatten off between 9
and 10~T. At $T=10$~K the increase in intensity has shifted up in
field to start at $\sim$8~T, and at $T=30$~K the intensity remains
constant with field. The increase in intensity with field at 1.8~K
is consistent with the expected spin-flop transition because once
the spins have rotated away from the $c$ direction magnetic Bragg
scattering is allowed at $(0,0,3)$ provided that the ordering
wavevector remains $(0,0,1)$.

With increasing temperature two effects are at work: (1) the field
at which the spin-flop transition occurs shifts up slowly with
$T$,\cite{luo-2004} and (2) the ordered magnetic moment $\mu$
decreases with $T$. Both effects would result in the reduction of
intensity with increasing temperature. However, the effect of (1)
is too small to explain the disappearance of the signal by 30~K
(see Luo {\it et al.} \cite{luo-2004}), so we deduce that the
signal disappears due to the reduction of the magnetic moment to
zero above the magnetic ordering temperature $T_m$.

In Fig.~\ref{fig:D3}b we plot the temperature dependence of the
scattering at ${\bf Q}=(0, 0, 3)$, in both zero applied field and
9.6~T. We confirm that the signal induced by application of the
magnetic field decreases to zero as the temperature is raised to
$\sim 20$~K.

\subsection{Analysis}

\begin{figure}[]
\begin{center}
\includegraphics{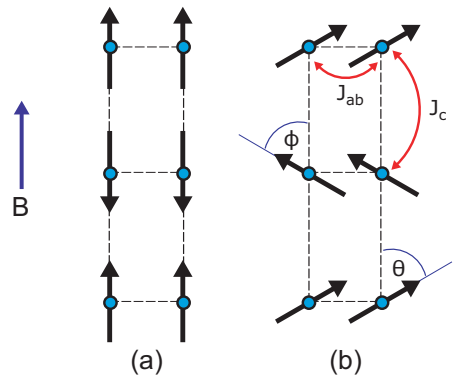}
\caption[Caption Caption]{(color online) (a) The ordered A-type
antiferromagnetic structure in the $a$--$c$ plane, with an
external magnetic field ${\bf B}$ applied parallel to the magnetic
moments. (b) Above a critical field $B_{sf}$ the system undergoes
a spin-flop transition, to a phase with spins at an angle
$\theta=\phi$ to the $c$-axis.
 } \label{fig:spin_flop}
\end{center}
\end{figure}

Figure~\ref{fig:spin_flop} shows the magnetic structure of the
low-field A-type antiferromagnetic (AF) phase (a), along with that
of the spin-flop (SF) phase (b). As discussed above, the large
increase in Bragg intensity at ${\bf Q}=(0,0,3)$ that occurs
between $\sim 6$~T and 9~T (Fig.~\ref{fig:D3}(a)) appears to
support the idea that a phase transition from an AF to a SF phase
occurs at this field. To connect the spin-flop transition with the
anisotropy gap described in Section II we extend the spin-wave
model to include an external magnetic field applied along the $c$
axis. We begin by calculating the effect of the field on the
spin-wave modes of the AF phase. This enables us to calculate the
critical field at which a phase transition would be expected to
occur. A term is added to the original Hamiltonian to represent a
vertical applied magnetic field, giving a new Hamiltonian:
\begin{equation}\label{new-hamiltonian}
\mathcal{H}'= \mathcal{H} + g \mu_B B \sum_{i} S_i^z \; ,
\end{equation}
where $\mathcal{H}$ is the original Hamiltonian
(Eqn.\ref{hamiltonian}), $B$ is the magnitude of the applied
magnetic field, and we assume $g=2$.

The spin-wave dispersion was derived from $\mathcal{H}'$ as
before. The field further splits the two modes, {\it i.e.} the
lower mode moves lower in energy, and the higher mode moves higher
in energy, as the field is increased. The magnitude
 of the gaps at the magnetic zone center
as a function of field, $B$, are then given by
\begin{eqnarray}\label{gap_AF}
    \hbar\omega_{gap}^{\pm}(B) &=&  2S\Big[\beta^2 + (2J_cD+D^2-E^2) \nonumber \\
    & & \pm 2\sqrt{\beta^2D(2J_c+D)+J_c^2E^2} \Big]^{\frac{1}{2}}
    \;,
\end{eqnarray}
where $\beta = g\mu_B B/2S$.

As the field is increased to a critical field, $B_{c1}$, a local
instability occurs when the energy of the lower mode falls to zero
and then becomes imaginary, and the system can no longer remain in
the low-field AF phase (Fig.~\ref{fig:spin_flop}a). The critical
field is determined by setting
$\hbar\omega_{gap}^{\pm}(B_{c1})=0$, resulting in the expression
\begin{equation}\label{Bc1}
    B_{c1} = \frac{2S}{g\mu_B} \sqrt{(D-E)(2J_{c} + D - E)} \; .
\end{equation}

$B_{c1}$ is the critical field at which we would expect the system
to `flop' out of the A-type antiferromagnetic phase
(Fig.~\ref{fig:spin_flop}a) into the spin-flop phase
(Fig.~\ref{fig:spin_flop}b) based on the closing of the gap with
increasing field.

A similar calculation can be performed by considering the
spin-wave modes in the SF phase, shown in
Fig.\ref{fig:spin_flop}(b), with $\theta=\phi$. In this phase
there are two modes at high field, and as the field is decreased
there exists a critical field $B_{c2}$ at which the lower mode
vanishes and the SF phase is no longer stable. At this field,
$B_{c2}$, the system returns to the AF phase. By calculating the
spin-wave dispersion in the SF phase we find the angle at which
the spins lie in the $a$--$c$ plane, as a function of field,
\begin{equation}\label{theta}
    \theta = \arccos \left( \frac{\mu B}{2S^2(2J_{c} - D + E)} \right)\;
    ,
\end{equation}
and an expression for the critical field, $B_{c2}$:
\begin{equation}\label{Bc2}
    B_{c2} = \frac{2S}{g\mu_B} \sqrt{\frac{(D-E)(2J_c-D-E)^2}{(2J_c+D-3E)}} \;
    .
\end{equation}

In a spin-flop transition $B_{c1}\geq B_{c2}$, so hysterisis is
possible. By evaluating Eqns.\ref{Bc1} and \ref{Bc2}, using the
values for the exchange and anisotropy constants determined by
fitting the inelastic data on Na$_{0.75}$CoO$_2$ at 1.5~K (Section
II C), we find values of $B_{c1}=8.19 \pm 0.2$~T and $B_{c1}=8.15
\pm 0.2$~T. Therefore the hysterisis is expected to be too small
to measure, and we take $B_{c}=8.2 \pm 0.2$~T as the critical
field. This value is in strong agreement with the experimental
data shown in Fig.~\ref{fig:D3}a. When the spins `flop' into the
SF phase we calculate $\theta = 87.8 \pm 0.5$~degrees (from
Eqn.~\ref{theta}), so the spins lie almost aniferromagnetically
perpendicular to the applied magnetic field. We note that a
mean-field approximation to this calculation also gives values
$B_{c}=8.2 \pm 0.2$~T and $\theta \approx 88$~degrees, and that
corrections to this calculation to account for the field offset of
7~degrees from vertical are small and within the given errors.

An estimate of the magnetic moment was derived from the
diffraction data by comparing the integrated intensities of a set
of nuclear reflections with the integrated intensity of the
(0,0,3) magnetic Bragg peak, using the A-type antiferromagnetic
model for the magnetic structure. From this we estimate a magnetic
moment of 0.11$\pm$0.07$\mu_{\rm B}$, in agreement with values
derived by other techniques. \cite{moment-reference} For this
calculation we used the free-ion form factor for Co$^{4+}$, which
may not be a good approximation given the metallic nature of
Na$_{0.85}$CoO$_2$.

\section{Discussion}

We have seen that the spin-wave dispersion does contain a clear
gap, and that at low temperature (1.5~K) it is well described by
our simple Hamiltonian extended to include two anisotropy terms
(Eqn. \ref{hamiltonian}). While including a term to describe a
uniaxial anisotropy ($D$) seems logical given that the spins do
lie along the $c$ axis, the need to introduce also an in-plane
anisotropy term, with parameter $E$ nearly two thirds as large as
the uniaxial parameter $D$, was unexpected. Although the form of
the in-plane anisotropy term may need more careful thought, we
have shown that in-plane anisotropy clearly exists in the
compound: the data cannot be described with the uniaxial
anisotropy alone. The magnitudes of the anisotropies $D$ and $E$
(0.096 and 0.059~meV respectively at 1.5~K) are very small in
comparison to the exchange parameters $J_c$ and $J_{ab}$ (12.2 and
6~meV). To refine the model further with a more realistic form for
the spin anisotropy would require a more detailed consideration of
the source of the anisotropy.

By fitting the same model to data measured at various temperatures
up to 24.4~K we have shown that both gaps disappear quite suddenly
at approximately 20~K. From the proximity of this temperature to
the bulk magnetic transition at $T_m \approx 22$~K found in
magnetization studies we infer that the spin waves have the same
origin as the $T_m$ transition. This may seem obvious, but the
strength of the spin-wave scattering in comparison to the small
size of the ordered moment made it important to confirm that the
observed spin excitations were really associated with the magnetic
order.

We have successfully observed the magnetic Bragg peak at ${\bf
Q}=(0,0,3)$ in Na$_{0.85}$CoO$_2$ with unpolarized neutrons by
inducing a spin-flop transition in a vertical magnetic field. The
transition occurred at a field of $B_{sf} \approx 8$~T, but was
broad, with a width of approximately 3~T. It is possible that the
broadening of the transition may be due to disorder in the
structure, which would lead to a spread of values of the
interlayer exchange constant $J_c$ and in turn lead to a range of
values for $B_{sf}$.

We have presented a simple calculation of the spin-flop transition
field that would be expected based on the anisotropy parameters
derived from the measurement of the spin gap in
Na$_{0.75}$CoO$_2$. The calculated value for this critical field
of $8.2\pm 0.2$~T is in strong agreement with the observed value
for Na$_{0.85}$CoO$_2$ of $B_{sf} \approx 8$~T. We take this as
further evidence that the A-type antiferromagnetic structure with
its associated gapped spin excitations is the same phase marked by
the transition temperature $T_m \approx $22~K in magnetization
data.

Although we have confirmed the magnetic structure and
characterized the excitation gap, the nature of the magnetic
ground state in Na$_x$CoO$_2$ is still in question. In an ionic
picture there are $(1-x)$ Co$^{4+}$ ions carrying spin
$S=\frac{1}{2}$ in a background of $x$ Co$^{3+}$ ions, which are
usually assumed to be non-magnetic. However, it is uncertain
whether these ions would be ordered or clustered or randomly
distributed, \cite{helme-2005} and there is also evidence that
Na$_x$CoO$_2$ is a good metal, which would suggest that an
itinerant picture might be more appropriate.\cite{itinerant} A
weakly itinerant ground state with strong spin fluctuations would
be consistent with both the small ordered moment ($\leq 0.2
\mu_B$) and the fact that the energy scale of the magnetic
excitations is much greater than the magnetic ordering
temperature.

In this vein, the clean gap we have observed in the spin-wave
dispersion  points towards a system of local moments with a small
symmetry-breaking anisotropy field, which is supported by our
successful description of the data using a simple spin-wave model
based on localized Co spins. However, the large intrinsic widths
are not consistent with a model of purely localized Co ions, and
can be taken as evidence of more metallic behavior. These features
of the magnetism of Na$_x$CoO$_2$ need to be taken into
consideration when assessing theories of the metallic state of
this material.

\section{Conclusions}

We have studied the low-energy region of the excitations in
Na$_{0.75}$CoO$_2$ and revealed that the excitations are gapped.
At low temperature the data were modelled best with a simple
spin-wave model with two non-degenerate modes, providing evidence
for both uniaxial and in-plane anisotropy in the compound. Gaps
from both modes were found to close at the bulk magnetization
temperature $T_m \approx 22$~K. We therefore conclude that the
spin waves observed in Na$_{0.75}$CoO$_2$ are associated with the
magnetic phase transition seen previously in magnetization data.

Furthermore, we have confirmed the A-type antiferromagnetic
structure underlying these excitations by performing neutron
diffraction measurements on Na$_{0.85}$CoO$_2$ in a large vertical
applied magnetic field. A spin-flop transition occurs at $\approx
8$~T allowing measurement of the magnetic Bragg peak at ${\bf
Q}=(0, 0, 3)$. The results support the assumption that the spins
lie along the $c$ axis. We have calculated the field needed to
overcome the anisotropy associated with the gap in the spin-wave
dispersion and shown that it is in good agreement with the
observed spin-flop transition field.

\begin{acknowledgements}
We would like to thank A. Hiess (ILL) for his help with the RESCAL
calculations. We would like to acknowledge the University of
Oxford and the Engineering and Physical Sciences Research Council
of Great Britain for financial support.
\end{acknowledgements}

\end{document}